\documentclass[twocolumn,trackchanges]{aastex701}

\usepackage{breqn}
\usepackage[caption=false]{subfig}
\usepackage{graphicx}
\usepackage[T1]{fontenc}
\usepackage{natbib}
\usepackage{amsmath}
\usepackage{float}
\usepackage{gensymb}

\usepackage[T1]{fontenc}

\DeclareRobustCommand{\VAN}[3]{#2}
\let\VANthebibliography\thebibliography
\def\thebibliography{\DeclareRobustCommand{\VAN}[3]{##3}\VANthebibliography}

\usepackage{graphicx}	
\usepackage{amsmath}	

\begin{document}

\title[PSR~J0901-4046 as an ultra-slowly rotating neutron star or a white dwarf?]{Exploring PSR~J0901-4046 as an ultra-slowly rotating neutron star versus a white dwarf }

\author[0000-0002-3020-9513]{Banibrata Mukhopadhyay}
\affiliation{Department of Physics, Indian Institute of Science, Bengaluru 560012, India}
\affiliation{Joint Astronomy Programme, Department of Physics, Indian Institute of Science, Bengaluru 560012, India}
\email{bm@iisc.ac.in}

\author[0000-0001-9829-7727]{Nirupam Roy}
\affiliation{Department of Physics, Indian Institute of Science, Bengaluru 560012, India}
\affiliation{Joint Astronomy Programme, Department of Physics, Indian Institute of Science, Bengaluru 560012, India}
\email{nroy@iisc.ac.in}

\author[0009-0000-6980-6334]{Zenia Zuraiq}
\affiliation{Department of Physics, Indian Institute of Science, Bengaluru 560012, India}
\email{zeniazuraiq@iisc.ac.in}

\author[0000-0003-4380-3033]{Chandrachur Chakraborty}
\affiliation{Manipal Centre for Natural Sciences, Manipal Academy of Higher Education, Manipal 576104, India}
\email{chandrachur.c@manipal.edu}

\author[0000-0003-0833-0533]{A. R. Rao}
\affiliation{Department of Astronomy and Astrophysics, Tata Institute of Fundamental Research, Colaba, Mumbai 400004, India}
\email{arrao@tifr.res.in}

\begin{abstract}
The source PSR~J0901-4046 shows a very long rotational period, of $\approx 75.88$ s. In fact, its long period suggests that it could be a white dwarf.
We examine the properties of the source and compare the outcomes under the hypotheses of the source being a highly magnetized white dwarf (B-WD) and neutron star.  We find that assuming it to be a B-WD with a surface field $\sim 10^{10}$ G explains many properties, including its storage/rotational energy, explaining luminosity. On the contrary, the neutron star-based idea, particularly 
if examined under the conventional spin down mechanism,
requires an unusually high surface magnetic field $\gtrsim 10^{14}$ G, even exceeding $10^{16}$ G, depending on the obliquity angle and emission mechanism. The source also sits very close to AR Sco, a white dwarf pulsar, in the period and period-derivative plane. We suggest that the source is a B-WD. Along with AR Sco, it being a B-WD constrains the death line model. B-WD hypothesis alleviates the necessity of an unusually high magnetic field hindering pair production for PSR J0901-4046 under the neutron star hypothesis. We argue that some rotation-powered pulsars with extreme properties may possibly be white dwarfs rather than neutron stars. 
\end{abstract}



\section{Introduction}

The radio pulsar PSR~J0901-4046 with its slow spin period of 75.88 s \citep{caleb} indicates that the neutron star, if it is the source of this pulsar, could be spinning at several orders of magnitude lower than its Keplerian (breakup) frequency. However,
typically, the spin period of radio pulsars ranges from 1.4 ms to 23.5 s,
further divided into various sub-classes, e.g. transient, millisecond
pulsar, magnetar (see, e.g., \citealt{caleb}). 
By tracking \citep{spindown0901} the exact arrival times of individual radio pulses over an extended time-frame with the uncertainty of rms arrival time $\sim 10^{-4}$ of the rotation period, the observed extremely small steady slow-down rate, $2.25\times 10^{-13}$ s s$^{-1}$, and without any wobble rule out the
period of PSR~J0901-4046 to be the binary orbital period. Therefore, the spin period of PSR~J0901-4046 alone is more consistent with a moderately
spinning white dwarf as the source of this pulsar. However, the deepest image of PSR~J0901-4046 shows
a diffuse shell-like structure appearing in its surroundings, possibly indicating an underlying supernova remnant \citep{caleb}. 
Nevertheless, the researchers
also noted that the surrounding interstellar field is highly complex and filled with diffuse emission. Therefore, 
multi-wavelength analysis is required to robustly confirm the definitive
origin of the shell.

It is evident in most neutron stars that the radio emission is expected to be 
strongly suppressed at the magnetic field above the quantum critical value, i.e. $4.414\times 10^{13}$ G \citep{BH1998}. However, the high magnetic field required to explain the radio emission from PSR~J0901-4046 as due to neutron star emission challenges this general belief.
This requirement of a high magnetic field also questions the nature of the spin evolution of neutron stars.
Moreover, the position of PSR~J0901-4046 in the $P-\dot{P}$ (where $P$ and $\dot{P}$ are the period and period derivative respectively) parameter space differs from the known ``magnetars'': soft-gamma repeaters (SGRs) and anomalous X-ray pulsars (AXPs), as well as the conventional neutron star population. Rather, it appears very close to AR Sco \citep{natphys}, which
is known to be a (highly) magnetized white dwarf and a synchronar, i.e. a binary whose emission is powered by synchronization between the components \citep{Katz17}.
Recent studies into its long-term evolution also pose differences in its behavior when compared to conventional magnetar observations \citep{spindown0901}. It may have other
differences in properties from conventional neutron stars, e.g., mass, apart from radius,
that we cannot yet access from observations. 
This, along with its low spin period, encourages us to scrutinize the actual nature of PSR~J0901-4046. In fact, there are other pulsars discovered with ultra-slow spin, e.g. 
GLEAM-X~J162759.5-523504.3 with a period $18.18$ min \citep{nature}. Although highly magnetized neutron stars are argued to have attained long rotational periods due to their bright coherent radio bursts \citep{Beniamini20}, the radio emission itself is expected to be suppressed at a strong field. Hence, this constrains
the magnetic field to an upper limit. Therefore, such ultra slowly spinning pulsars can be magnetized white dwarfs, as suggested by \cite{katzApSS} for  GLEAM-X~J162759.5-523504.3. 
The lower magnetic field of a white dwarf can be compensated by its bigger size (see equation (\ref{multipole}) below), leading to the available energy to be same, if not more, as of a neutron star.

Indeed, there has been a substantial amount of progress in both the discovery of new ultra-slowly spinning pulsars and our understanding of the objects that produce them. For instance, many of them are consistent with being white dwarfs with M-dwarf companions. The radio transient GLEAM-X J0704-37, having a period of 2.9 hr has been discovered to have an optical counterpart, consistent with a cool main-sequence star of spectral type M3 \citep{hurley}. The radio transient ILT J1101+5521 establishes that long-periodicity like 125.5 min originates from orbital motion modulating the observed emission \citep{ruiter}. Being a polar system, magnetic interaction has synchronized the rotational and orbital periods of ILT J1101+5521. Another source is J1634+44 of 
period $\sim 14$ min, which exhibits two pulses produced every five periods with a waiting time of two or three periods \citep{bloot}. This is expected to be a result of spin-orbit coupling in a binary system, where the expected companion being an ultracool dwarf or another white dwarf induces beamed radio emission on the white dwarf.
In addition, a very long period ($\sim 22$ min) pulsar GPM J1839-10 can be explained by binary interactions \citep{horvath1}. The authors showed that the radio emission arises by the intersection of magnetic axis of a rotating white dwarf with the wind from its companion. This explains the peculiar pulse modulation. The model further reproduces the emission profile and geometry of J1912-44. 
\cite{horvath2} also examined a 44 min period pulsar, ASKAP/DART~J1832-0911, by its evolution of the pulse morphology and the rotation measure variation, and found it to be a nearly synchronised close binary system, like AR Sco though with much higher luminosity.

There has been a proposal of the existence of highly magnetized white dwarfs (B-WDs)
with mass possibly exceeding the Chandrasekhar limit of $1.4M_\odot$. In a series of papers (see, e.g., \citealt{DM12,DB2013,DM15,SSM2015,BM2017,KM2019,KMMB2020,KMTBM2020,BHGTM2022,deb21,ZMStars}), via theory and simulation, the authors showed that magnetized, rotating white dwarfs can have mass even exceeding $2.5M_\odot$, corroborating with the observational inference (e.g. \citealt{Howell,scalzo10}). However, rotating B-WDs or moderately magnetized white dwarfs have been hardly explored as a radio pulsar source. Can they be the candidates for PSR~J0901-4046 and even 
GLEAM-X~J162759.5-523504.3, particularly when there is no signature of accretion (X-ray) and no constraint on binary modulation? Earlier authors, by theory and simulation \citep{BM2017,ZMStars}, showed that by accretion, a low mass sub-Chandrasekhar, relatively lowly magnetized white dwarf could lead to a highly magnetized, super-Chandrasekhar white dwarf. Also, 
the same group showed that some SGRs/AXPs might be B-WDs \citep{BR2016}, which are inferred to exhibit relatively low magnetic fields. All these works suggest that some magnetized
compact stars and pulsars could fit for white dwarfs as well; and may even be better than neutron stars in terms of magnetic field budget. This is often useful as it could avoid unusually high magnetic fields as the magnetar model needs, yet the object could have a good reservoir of the magnetic field due to its much larger size \citep{multipole2}. 

From the above discussion, it is apparent that some rotation-powered pulsars showing extreme observed properties could as well be white dwarfs, apart from neutron stars, as has been shown before (e.g. AR Sco). This paper takes the case of PSR~J0901-4046 explicitly and succinctly summarizes the energy generation and evolution calculations to show that PSR~J0901-4046 could be harboring a B-WD. Additionally, we explore the `death line', specifically in the context of white dwarfs. This demonstrates that the B-WD hypothesis alleviates the requirement of an unusually high magnetic field required for PSR~J0901-4046 under the neutron star hypothesis.

In an event of the absence of magnetospheric
pair production, weak plasma loading from the interstellar medium (ISM) may provide adequate magnetospheric electrons to support coherent radio emission by electron cyclotron maser emission (ECME) in ultra-long-period neutron stars such as PSR~J0901-4046 at much lower magnetic fields \citep{Lillia}. This process does not require accretion and/or related interaction from a binary companion, as is absent currently in PSR~J0901-4046 \citep{caleb,Ronchi_2022}. In such a scenario, the source would resemble an auroral magnetosphere rather than a conventional accretion-powered pulsar. However, the concern remains whether the available rotational or magnetic energy can sustain the required particle acceleration and coherent emission efficiency.

Hence, in the present work, we suggest that PSR~J0901-4046 is possibly a B-WD.
We argue that it had a binary partner, which is exhausted now \citep{Hillman_2016}, which supplied mass by accretion, leading to its evolution to a B-WD from a conventional lowly magnetized white dwarf \citep{BM2017,ZMStars}. Also, in place of complete exhaustion, the companion may have a mass too small to be detected \citep{parsons2026,belloni2023}. Moreover, the origin of this B-WD could be by the merger of two white dwarfs which, hence, is not expected to leave any companion today (see \citealt{dwd_merger} for the exploration of the white dwarf source J2211+1136 under this formation scenario, having a very similar rotation period $P = 70.32$ s). 

Thus, in the absence of a confirmed observation of a binary partner at present, PSR~J0901-4046 is suggestive of being the isolated white dwarf pulsar, with pulsation driven by magnetospheric processes as suggested in other transient sources \citep{GilZhang05}. This is unlike AR Sco where the emission is thought to arise in connection to binary interactions \citep{natphys,Katz17}. In the next section, we revisit the energy generation and evolution calculations of rotating compact objects. We consider primarily dipolar electromagnetic emission with, 
however, an example case 
of quadrupolar emission, with
the braking index more than 3. We also outline the `death line' under different models prevalent in the literature, specifically in the context of B-WDs. In Section 3, the possible evolutionary scenario of such objects is discussed, particularly in the context of PSR~J0901-4046, and the locations of  PSR~J0901-4046 and other similar objects in the $P-\dot{P}$ plane are examined, juxtaposing them with the current ideas of death lines. The last section (Section 4) summarizes the present work with concluding remarks.

\section{Modelling pulsar as a multipolar object}
\label{sec:model}

It is quite common to assume a rotating magnetized neutron star or a white dwarf, in general compact stars, to be a rotating dipole \citep{Goldreich69}. 
Then, from standard electromagnetic theory (e.g. \citealt{jackson}) and assuming that the dipole nature of the star drives the rate of rotational energy released, we can write for a star of moment of inertia $I$, angular velocity $\Omega$,
magnetic field at pole $B$, obliquity angle $\alpha$, radius $R$ \citep{PETRI2015},
\begin{eqnarray}
I\Omega\dot{\Omega}=-\frac{4\Omega^4\sin^2\alpha B^2 R^6}{3c^3},
\end{eqnarray}
where $c$ is the speed of light. Over the years, many 
neutron stars, e.g., Crab pulsar \citep{Tong_Kou2015}, SGR 1806220 \citep{Harding_1999}, and some white dwarf pulsars, e.g.,
AR Sco \citep{BuckleyNatA}, have been modeled as a rotating dipole. 

If we assume that the magnetic field effect of the compact star 
is more complicated than that based on the simpler dipole assumption \citep{PETRI2015,AK2002}, which seems to be more natural for a significantly extended rotating body like a white dwarf \citep{multipole1,multipole2}, 
then 
the effective electromagnetic energy released 
rate is \citep{PetriGR13,PETRI2015,PetriQED16}
\begin{eqnarray}
   \dot{E}_{em}\propto 
   -\Omega^{m+1}B^2R^{m+3}\sin^2\alpha.  
\label{multipole}
\end{eqnarray}
Moreover, observations argue $\dot{\Omega}\propto\Omega^n$ with the braking index $n\neq 3$ \citep{braking1,braking2,Parthasarathy_BrakingIndex}. 
When we have a 100\% equivalence between the rotational and electromagnetic energies in the form of the rotating idealistic simple dipole model, 
we have $m = n = 3$. 
In practice, the field may have 
non-dipolar components ($m\neq 3$) and/or there may be more mechanisms at play other than dipolar braking ($n\neq 3$), including extended magnetosphere, particularly for a bigger
compact object like a white dwarf with a complicated magnetic structure. Therefore, the effective indices, controlling the rate of magnetic energy release (by the effective magnetic field strength and magnetosphere) and rotational energy, deviate from $3$. See the work of \cite{Petri15b,Petri17,PetriQED16} showing how the general expansions for an extended body with magnetosphere for multipolar/complicated fields leading to the power are determined.

Now assuming the emission to be purely from electromagnetic process with $m=3$ for dipolar and $m\neq 3$ for non-dipolar fields, by  equating $I\Omega\dot{\Omega}$ to equation (\ref{multipole}), we obtain
\begin{equation}
\dot{\Omega}=\tilde{K}B^2\Omega^m,
\label{multipoleB}
\end{equation}
where $\tilde{K}=K^\prime\sin^2\alpha R^{m+3}/I$ with $K^\prime$ being the proportionality constant in equation (\ref{multipole}) which varies with $m$; for $m=3$, $K^\prime=4/3c^3$.

Realistically, effects other than electromagnetism, like, pulsar winds, gravitational radiations, etc., could also extract
energy from a pulsar. This leads to
\begin{eqnarray}
\dot{\Omega}=K\Omega^n,
	\label{omn}
\end{eqnarray}
where $n\neq m$ in general where the emission is by electromagnetic as well as non-electromagnetic mechanisms,
and $K$ is a constant. 
Considering equations (\ref{multipoleB}) and (\ref{omn}) to be the same, we obtain the magnetic field 
\begin{eqnarray}
	B=\sqrt{\frac{K\Omega^{n-m}}{\tilde{K}}}.
\label{multipoleB2}
\end{eqnarray}

On the whole, the above proposal captures dipolar, non-dipolar and non-electromagnetic emissions. For $m=n=3$, the source is dipolar and purely electromagnetic; $m=n\neq 3$ corresponds to a non-dipolar but purely electromagnetic source; $m\neq n\neq 3$ corresponds to a non-dipolar source with electromagnetic and non-electromagnetic emissions; $n\neq m=3$ corresponds to a dipolar source with electromagnetic and non-electromagnetic emissions.

Now by expressing equation (\ref{omn}) in terms of $P=2\pi/\Omega$, we obtain at an 
arbitrary time
\begin{eqnarray}
	t_a=\frac{P}{(n-1)\dot{P}},	
	\label{const}
\end{eqnarray}
	when initial time period is assumed to be negligibly small.

\subsection{Death line}

For any pulsar, the death line in the two-dimensional pulsar parameter plane of 
$P-\dot{P}$ (or even $P-B_s$ and $P-\Phi$, where $B_s$ and $\Phi$ are, respectively, surface magnetic field and polar cap potential) is the line which separates the parameter region that enables the production of radio emission from the one that cannot.
It was conjectured \citep{Sturrock} that copious pair production in the inner magnetosphere is responsible for radio emission. However, the exact death line is model-dependent,
and further depends on the obliquity angle and the equation of state. Two of the popular models for the polar cap gap are \citep{bing}:
(1) vacuum gap, (2) space-charge-limited flow. Each model further can give two possible
death lines, depending on the origin of $\gamma$-ray photons, whether they are produced by
curvature radiation (CR) or inverse Compton scattering (ICS). For the vacuum gap model,
the death lines for CR and ICS for multipolar magnetic field configuration are,
respectively,
\begin{eqnarray}
	\log\dot{P}=\frac{9}{4}\log P-16.58+\log(R_6)
	\label{vac-cr}
\end{eqnarray}
and
\begin{eqnarray}
	\log\dot{P}=-\frac{2}{11}\log P-14.5+\frac{8}{11}\log(R_6),
	\label{vac-ics}
\end{eqnarray}
where $R_6$ is the stellar radius in units of $10^6$ cm. For the space-charge-limited
flow model, the respective death lines are
\begin{eqnarray}
	\log\dot{P}=2\log P-16.52+\log(R_6)
	\label{vac-cr}
\end{eqnarray}
and
\begin{eqnarray}
	\log\dot{P}=-\frac{7}{11}\log P-16.79+\frac{8}{11}\log(R_6).
	\label{vac-ics}
\end{eqnarray}
Note that depending on the exact physics involved, the coefficients in various terms of the death line may slightly vary, without however affecting the gross physics. Hence, the same/similar death lines are considered, independent of the nature of compact objects, for all practical purposes \citep{GilZhang05,CR1993}.
Further, there are other uncertainties in the death line model apart from the magnetic field topology and pair production mechanism, e.g., pulsar obliquity. This can lead to an overall death valley, as opposed to a single death line \citep{SM2023}.

\section{Evolution of a star by accretion}

Following previous work \citep{BM2017}, we consider the evolution of a compact star by 
accretion based on the conservation of momentum and magnetic flux. We do not
repeat the details here. In summary, due to accretion, the mass of the compact
star, particularly a white dwarf, increases, the radius decreases, and the magnetic field increases. Also, its spin
is coupled to the evolutions of mass, radius, and magnetic field. 

We, however, consider the following simplifications:\\
(1) We do not include any physics related to cataclysmic variables.\\
(2) Consequently, mass eruption/outflow, nuclear instability, possible variation of accretion rate during evolution have been assumed to be unimportant.\\
(3) We do not explicitly solve the coupled Euler-Maxwell equations to capture our evolution.\\
(4) We do not intend to embark into detailed emission properties, including highly structured radio pulses, without optical and X-ray,  and constraints on polarimetric signatures.

\begin{figure*}
	\includegraphics[width=9.5cm,page=1]{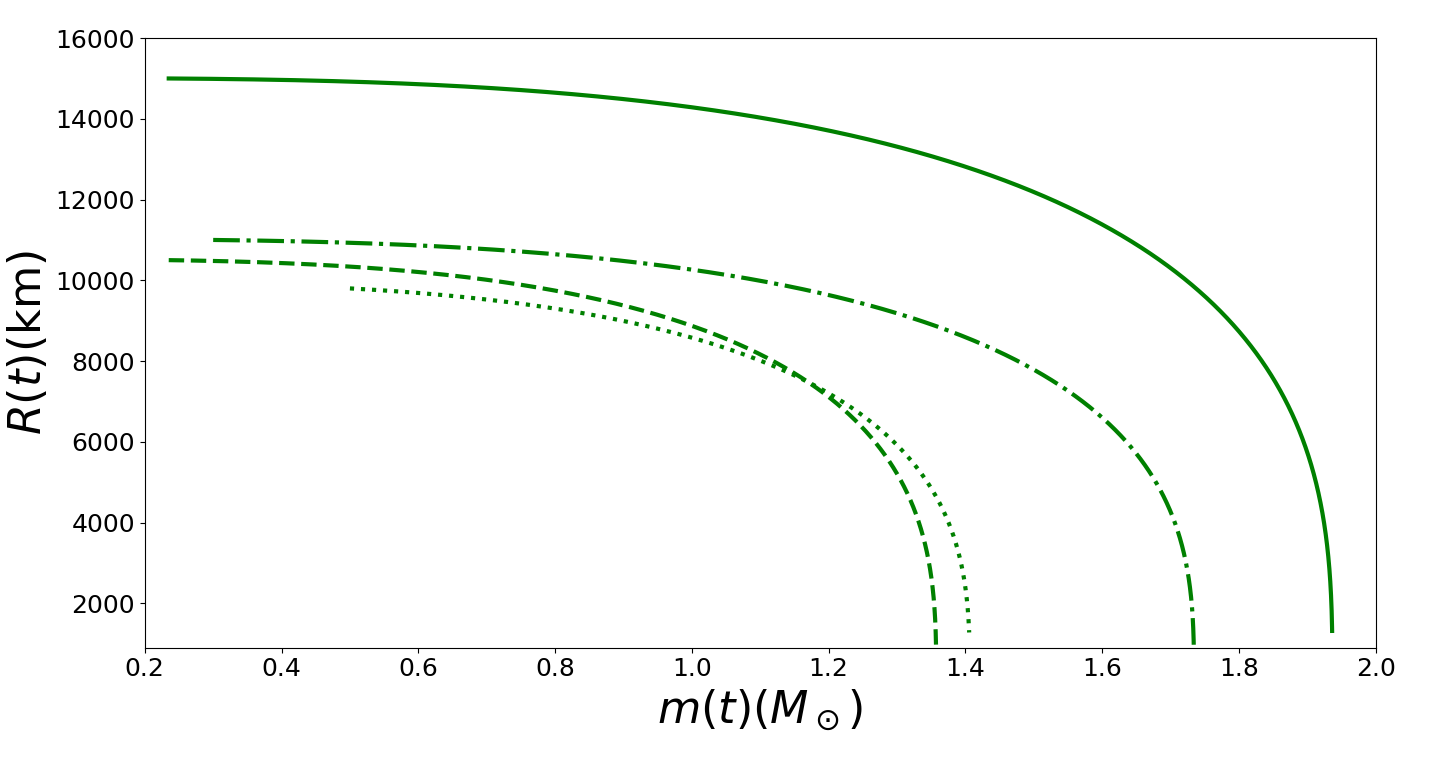}
	\includegraphics[width=9.5cm,page=1]{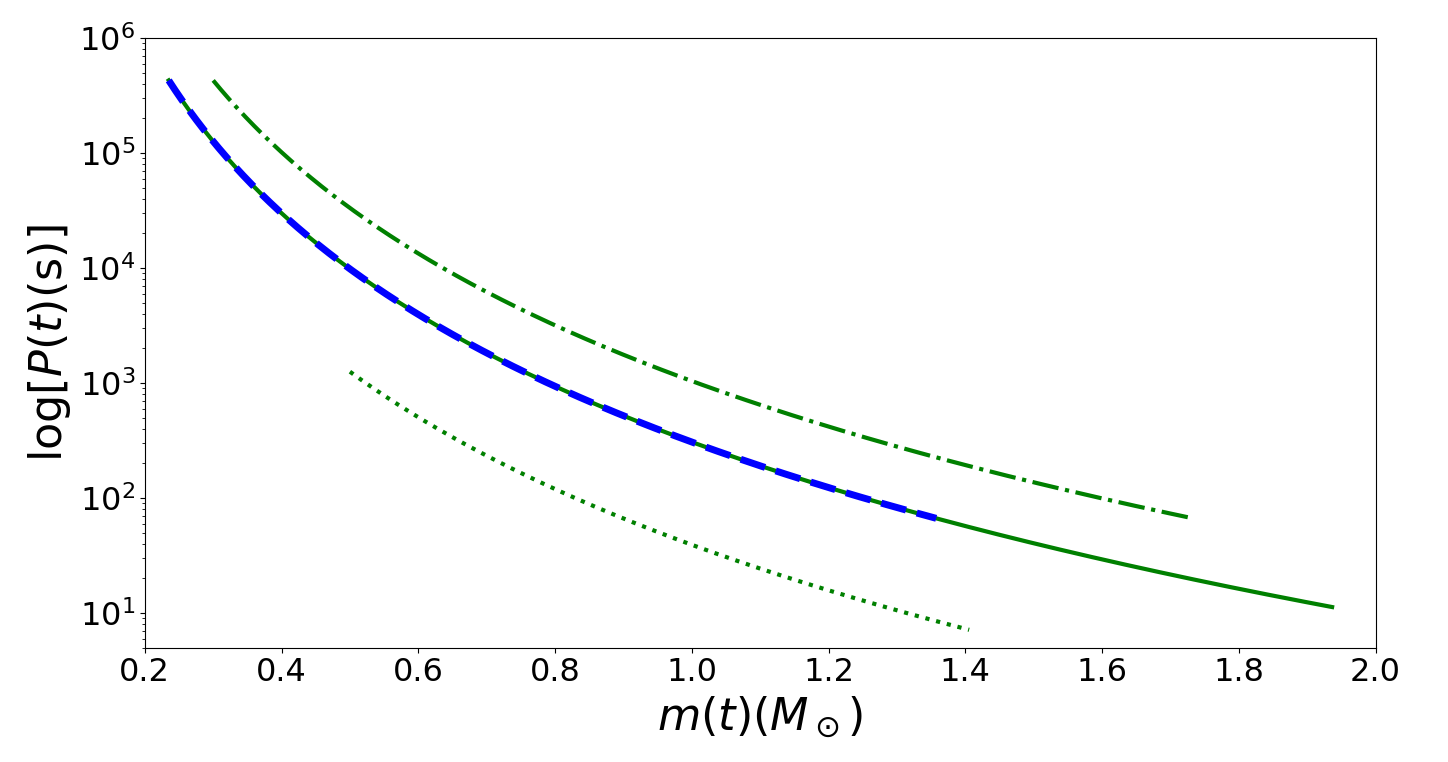}
	\includegraphics[width=9.5cm,page=1]{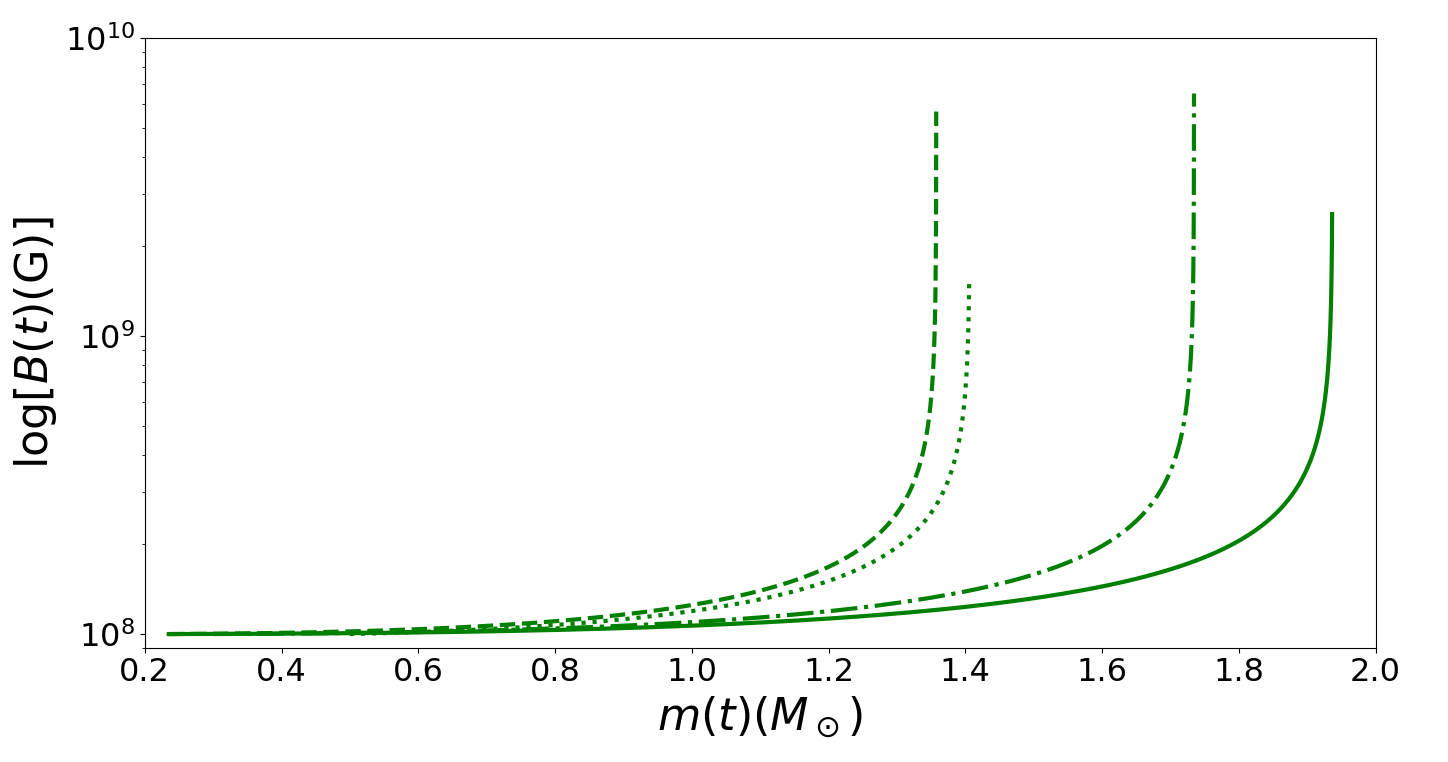}
	\caption{Evolution of the parameters of representative white dwarfs (radius - top left panel; rotation period - top right panel; surface magnetic field - bottom panel) by accretion for various initial sets of mass, radius, magnetic field, and spin, when the accretion rate is $3\times 10^{-9}M_\odot$/Yr. Note that the inner magnetic field also evolves and its final value ranges $10^{13}-10^{14}$ G. The time goes from left to right as mass grows. The stars representing the solid and dashed lines in the $P$-plot coincide during their respective evolutions, because they start with the same mass, radius and spin. The dashed curve terminates earlier due to its different net initial pressure.  
}
    \label{evol-acc}
\end{figure*}

Some of the above simplifications are relaxed by \cite{ZMStars} and in a companion paper by Zuraiq \& Mukhopadhyay (hereinafter Paper II: ``Exploring the origin of slow-rotating pulsars: Emission from STARS simulated
white dwarfs''). 
Throughout, we restrict the central density to $\lesssim 10^{10}$ g/cc, which consequently assures stability against electron capture \citep{VM2014,ChamelPRD2015,gupta2020}. Further, the inner magnetic field is restricted to $\sim 10^{14}$ G, which allows us to ignore the Landau quantization effect \citep{udas_JCAP2014,SSM2015}.
During evolution, once the field
goes above a certain value, it inhibits accretion, and the star starts
emitting radiation as a pulsar till it exhausts significant magneto-rotational energy, when, subsequently, it again starts accreting matter.
Fig. \ref{evol-acc} shows, based on a representative simplistic model \citep{BM2017}, how low-mass white dwarfs evolve by accretion and become
massive, and can even become a super-Chandrasekhar white dwarf at the end of accretion
phase when the surface (and hence interior) magnetic field becomes very high. We consider four sets of initial mass, radius, magnetic field, and spin of white dwarfs under accretion and depict their respective evolution. It is clear that the final evolved white dwarf depends on the properties of the initial white dwarf. The evolution stops when the magnetic field becomes too high to hinder any accretion due to, e.g., propeller effects \citep{ghosh-lamb}.
We explore a detailed time-dependent numerical simulation capturing the full-fledged evolutions of magnetized main sequence stars to magnetized white dwarfs and even B-WDs in the  companion work, Paper II. See below for preliminary results.

Figure \ref{wd90} shows how a $1.29M_\odot$ orthogonally rotating white dwarf behaving as a rotating dipole,
formed at the end of accretion-based evolution shown in Fig. \ref{evol-acc}, further evolves as a pulsar. We represent it as the evolution of constant surface magnetic field (Fig. \ref{wd90} top left panel), constant
luminosity (Fig. \ref{wd90} top right panel) and constant age (Fig. \ref{wd90} bottom panel) 
$P-\dot{P}$ lines. We place it with reference to the observed pulsars and the model death lines for a white dwarf. It also shows the possible past and
future trajectories of PSR~J0901-4046 in the $P-\dot{P}$ plane (dashed line) based on equation (\ref{omn}). 
Note that due to $n=4\neq m$ the trajectory of PSR~J0901-4046 is not parallel to the constant magnetic field lines, where its  current surface field turns out to be $\sim 10^{10}$ G, as given in Table \ref{table:field}, with orders of magnitude higher inner field \citep{SSM2015,deb22,UnivZD}.
The chosen parameters for Fig. \ref{wd90}, and other
subsequent figures and Table \ref{table:field}, corroborate with physically sensible evolutions and emission of pulsars based on 
rigorous simulations (e.g. by STARS, reported below and in the companion Paper II). Although there are certain quantitative differences, e.g. the exact matching of the set of the radius and magnetic field of simulated stars of a given mass, they qualitatively support each other.

Figure \ref{wd9} presents the same as above,
except for a super-Chandrasekhar white dwarf pulsar with $n=6$. 
Note that the evolutions of both super-Chandrasekhar and sub-Chandrasekhar cases are very similar/same (with a marginal shift in the constant magnetic field lines). A smaller $\alpha$ would simply scale up the surface magnetic field, keeping other properties similar as long as there is no significant change in the central magnetic field and other properties kept unchanged. Nevertheless, white dwarfs with smaller $\alpha$ with similar surface fields are possible with different field geometry and internal fields leading to different radii, as reported in Table~\ref{table:field}.
The table also confirms that 
depending on the magnetic field geometry, the surface magnetic field could be lower, even $\lesssim 10^9$ G, for higher $\alpha$.

However, all of the above properties argue that not all death line models favor radio observations 
of PSR~J0901-4046 as a white dwarf. Importantly, had we considered the pulsar and the model death lines for a neutron star of size
13 km, still CR based death lines are unable or at best cannot guarantee for radio observations, as shown in Fig. \ref{ns-p}.
However, PSR~J0901-4046 being a neutron star argues for the possible birth of slowly spinning highly magnetized neutron stars; even due to accretion, similar spin remains as those in the cases shown in Fig. \ref{evol-acc} for white dwarfs. If this is so, it is a bit unusual, as accretion is expected to increase spin unless it was born with an extremely slow spin. While magnetars, and hence AXPs/SGRs, are also slowly rotating with $P\sim 10$ s (hence, their frequency is several factors bigger than PSR~J0901-4046), their magnetic fields $\gtrsim 10^{13}$ G and they exhibit other activities, e.g. in X-rays. However, PSR~J0901-4046 is active only in radio, even though it, being a neutron star as above, suggests the field to be $\sim 10^{14}$ G (see Table \ref{table:field}). This argues it to be different from AXPs/SGRs. Indeed, in the $P-\dot{P}$ plane, magnetars are not close to PSR-J0901-4046.
For $\alpha<90^\circ$, the magnetic field
may go well above Schwinger limit, posing a further problem to explain radio emission from a neutron star based model, while for white dwarfs the field has never been too extreme due to their much larger size.

\begin{table}[!h]
\centering
\begin{tabular}{llllllll}
\hline
 Geometry &       $M/M_\odot$ & $R/\rm{km}$ & $\alpha$ & $n$ & $m$   & $B_s$/G            \\ \hline
 Dipole & 1.29         & 1031.89  &   $90^\circ$  & $4$ & $3$ & $1.7 \times 10^{10}$ \\
 Dipole & 1.32 & 4453.42 & $2^\circ$ & 6 & 3 & $3.2\times 10^{10}$ \\
 Dipole & 1.93 & 1298.84 & $90^\circ$ & $6$ & $3$               & $1.7 \times 10^{10}$ \\
 Dipole & 1.69 & 4712.23 & $2^\circ$ & 6 & 3 & $3.2\times 10^{10}$\\
 Dipole & 1.93 & 13 & $90^\circ$ & $6$ & $3$ & $1.7\times 10^{14}$\\
 Quadrupole & 1.93 & 13 & $90^\circ$ &  6 & 5 & $2.4\times 10^{19}$\\
 Quadrupole & 1.93 & 1298.84 & $90^\circ$ & 6 & 5 & $2.4\times 10^{13}$\\
\hline
\end{tabular}
\caption{Magnetic fields obtained in various cases considered including those in Figs. \ref{wd90}, \ref{wd9}, \ref{ns-p} and \ref{quadNW} for J0901-4046. See other parameters in the respective figure captions.}
\label{table:field}
\end{table}

Hence, the CR-based death lines appear to be ruled out for both the types of compact objects. For white dwarfs, this is due to their lower magnetic fields and, hence, their larger radius of curvature, leading to smaller CR energies, insufficient for pair production.
Even though for neutron stars CR energies improve due to stronger fields and much lower radius, stars are not well above the CR death lines.
For the ICS based pair production to work successfully, the temperature is an important factor. However, the resonant ICS is still not energetic enough to explain 
pair-production, due to lower surface fields. Nevertheless, white dwarfs with relatively low surface temperature compared to neutron stars are still expected to exhibit non-resonant ICS for pair creation and consequently pulsar emission. Note that the death lines shown in the present work are for resonant ICS; however, non-resonant ICS death lines are expected to be even lower in the parameter space. In that sense, the resonant ICS death lines are a stricter limit on pulsar emission, which have been satisfied by our current model. Previous authors \citep{GilZhang05,Zhang1997} argued for the possibility of coherent emission mechanism from a distant source without companion by non-resonant ICS for pair creation, particularly with extreme magnetosphere. 
We suggest similar complicated
non-resonant ICS processes to take place in PSR~J0901-4046 
with higher braking index. This is more so for B-WDs when the possibility of multipolar fields/magnetosphere effects is more open (see below). 

An additional mechanism that could explain pulsar emission from isolated compact sources, apart from magnetospheric pair production, is ECME. Initially explored for binary sources with interactions \citep{Zhang_ECME}, recent work \citep{Lillia} has highlighted how isolated white dwarfs and neutron stars, accreting from ISM plasma, may result in ECME. There are a few caveats to consider here however. ECME is favored for low fields. For neutron stars of surface field $B_s \gtrsim 10^{12}$ G, the star enters the propeller regime \citep{Lillia} and accretion is disfavored. Further, the compact star must be sufficiently slow-moving and embedded in regions with dense enough ISM. It is unclear if PSR~J0901-4046 meets these requirements. Accretion of the ISM material is expected to additionally produce soft X-rays, which are not observed from PSR~J0901-4046 yet (however, the X-rays may be below the current detection threshold). Nevertheless, keeping the above caveats in mind, note that PSR~J0901-4046 is $\sim 300-400$ pc away and, hence, ECME from it being an isolated B-WD may not be ruled out, especially if the field is additionally supported by multipolar components.  
 
Note that the constant magnetic field and constant luminosity lines, respectively, shift down and up for neutron star w.r.to white dwarf based cases. In other words, for the same $P-\dot{P}$ set, the magnetic fields for a neutron star are higher and luminosities are lower.

\subsection{Possible quadrupolar emission}
Above analyses assume the star to be emitting dipole radiation. It is, however, quite a simplified assumption. Compact stars, particularly the white dwarfs, which are bigger in size, are much more complicated emitters including complicated geometry with extended magnetosphere. Hence, there is no reason for $m=3$ to be strictly valid. Fig. \ref{quadNW} shows how a $1.93M_\odot$ orthogonally rotating compact object with $n=6$ and $m=5$ behaving as the evolution of the constant magnetic field $P-\dot{P}$ lines. It shows that while the neutron star based model suggests the magnetic field to be too high and above $10^{19}$ G, hence absurd, for white dwarf it is $\sim 10^{13}$ G. Although the white dwarf based model field is higher compared to the fields inferred from conventional white dwarfs, it is still below the critical field, as given in Table \ref{table:field}. It is possible that higher quadrupolar field is localized and is not visible/observable on larger scales.
In reality $m$ may be between $3$ and $5$ and, hence, the magnetic field varies accordingly.

The above analysis argues that nothing goes wrong with the consideration of PSR J0901-4046 
to be a white dwarf, either a rotating dipole or quadrupole; either sub-Chandrasekhar but magnetized or B-WD. 
While apparently, the idea of a neutron star is not ruled out, it has to be very highly magnetized when we hardly witness such a slowly spinning neutron star. Either simple dipole or with additional effects including higher order non-dipolar components, 
for a neutron star the field of PSR~J0901-4046 could range in $\sim 10^{14}-10^{19}$ G. This is difficult to imagine (and seems to be absurd for quadrupole) when the source has no activities other than radio.

\subsection{Exploration for  GLEAM-X J162759.5-523504.3}
Our exploration with white dwarf and B-WD is expected to be equally successful for the extremely slowly rotating (with period 1091 s) GLEAM-X~J162759.5-523504.3. Considering it to be a neutron star with dipole assumption, its magnetic field turns out to be unusually as high as $\sim 5\times 10^{16}$ G for $\alpha=90^\circ$, which is even much higher than the field of usual magnetars, i.e. SGRs/AXPs (and will be further higher for 
$\alpha<90^\circ$). The underlying rotational energy is also very low, well below its observed luminosity. However, under the B-WD hypothesis, with the parameters same as in Fig. \ref{wd9}, except $K=0.084~{\rm CGS}$, the surface magnetic field turns out to be $\sim 10^{10}$ G with much higher inner fields, 
which is quite reasonable for a B-WD, and rotational energy $\gtrsim 10^{33}$ erg/s, which could explain the luminosities of GLEAM-X~J162759.5-523504.3 in radio band.  
Hence, we suggest GLEAM-X~J162759.5-523504.3 is most likely to be a B-WD. In Paper II, we explore this source, along with others, in detail.

\subsection{STARS simulations}
We have initiated, through a detailed time-dependent numerical simulation, the full-fledged evolution of magnetized main sequence stars to B-WDs \citep{zz_stars,ZMStars} using the Cambridge stellar evolution code STARS \citep{eggleton}. The details of the simulation and stellar evolution calculations are given in Paper II.

To verify the current results by rigorous computations, we  obtain preliminary results; also reported in detail in the Paper II, required to reproduce the magnetic field that explains $P$ and $\dot{P}$ observed for PSR~J0901-4046. Some results 
are listed in Table \ref{table:STARS_WDP}. Note here that the value of the current surface magnetic field does not depend on $n$ - it only affects the evolution of the field in time. More detailed exploration of the parameter space and extension to more sources is given in Paper II.
Table \ref{table:STARS_WDP} shows the results based on full-fledged Cambridge STARS simulation leading to stellar evolution with realistic white dwarf properties corroborating with our electromagnetic energy released. 
We examine those cases where the simulated white dwarf's $B_s$, $M$ and $R$ are in agreement with the prediction from the dipole model. We also consider different magnetic properties (e.g. profile) for the evolution, leading to diversity of magnetic fields in the white dwarfs. This is explained in detail in Paper II.

\begin{figure*}
 	\includegraphics[width=7.5cm,page=1]{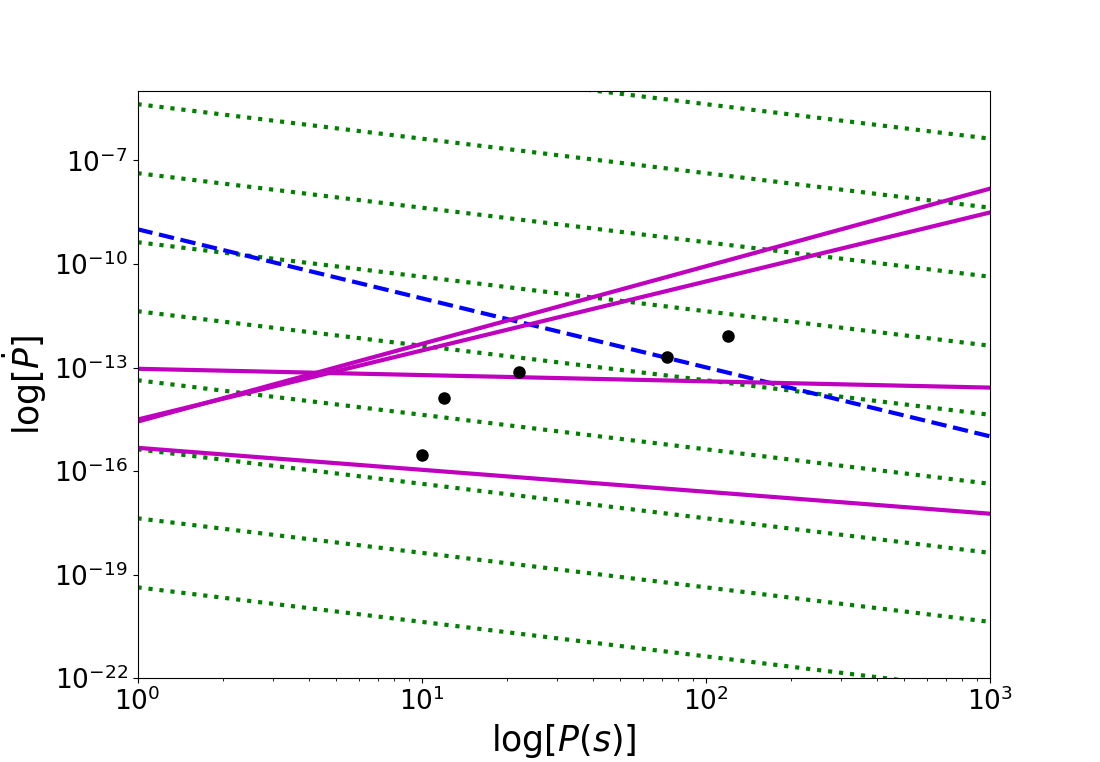}
	\includegraphics[width=7.5cm,page=2]{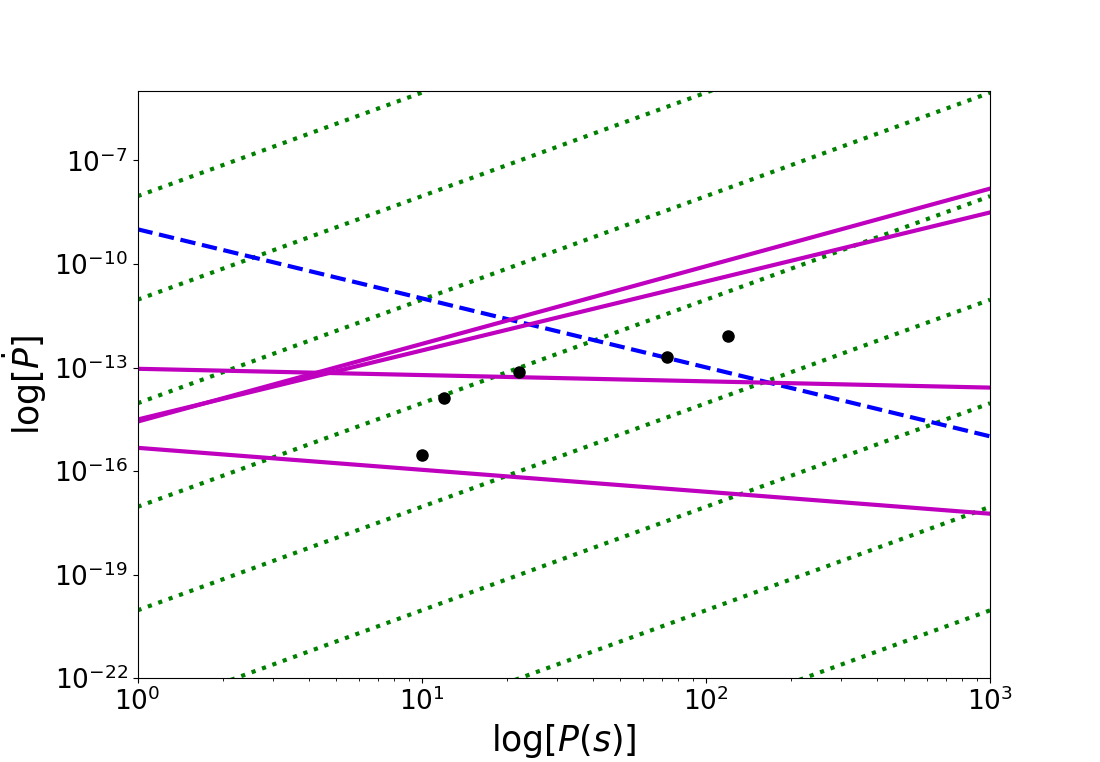}
	\includegraphics[width=7.5cm,page=3]{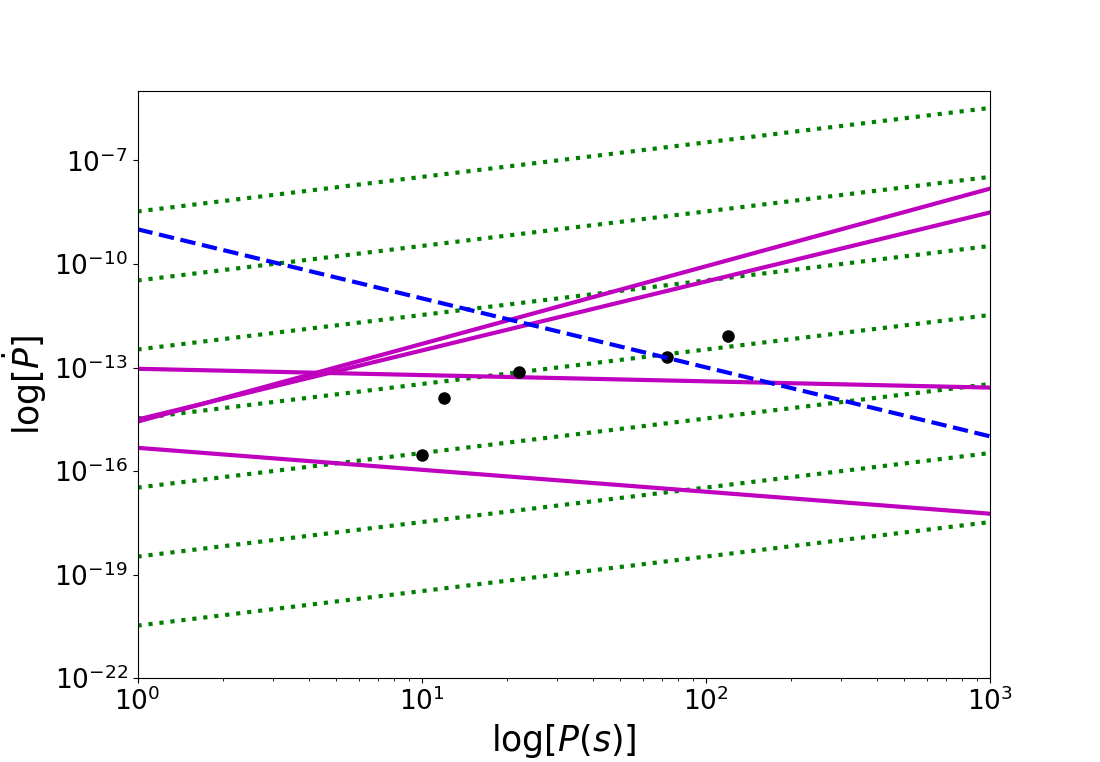}
	\caption{
    Evolution of a white dwarf of mass $1.29M_\odot$, radius $1031.89$ km and obliquity
	angle $\alpha=90^\circ$ with four
	model death lines for constant surface magnetic field (top left panel), constant luminosity (top right panel),
	and constant age (bottom panel). 
	Other parameters are: $n=4$, $K^\prime=4/3c^3$,
	$K=8.2\times 10^{-8}$ CGS. The various constant surface magnetic field lines from
	top to bottom correspond to $10^{14}$ to $10^6$ G with a common ratio 
	between two successive values 10. The various constant luminosity lines
	from top to bottom correspond to $10^{43}$ to $10^{22}$ erg/s with a 
	common ratio $10^{3}$. Finally, the various constant age lines
	from top to bottom correspond to $3.33$ to $3.33\times 10^{12}$ Yr with a common ratio $0.01$. 
	The filled circles from right to left are observed data for AR~Scorpii, J0901-4046, J0250+5884, J2251-3711, J2144-3933. The dashed line passing 
	through J0901-4046 corresponds to equation (\ref{omn}).
    The surface magnetic field for PSR~J0901-4046, according to equation (\ref{multipoleB2}), turns out to be $\sim 10^{10}$ G.
}
    \label{wd90}
\end{figure*}

\begin{figure*}
    \includegraphics[width=7.5cm,page=1]{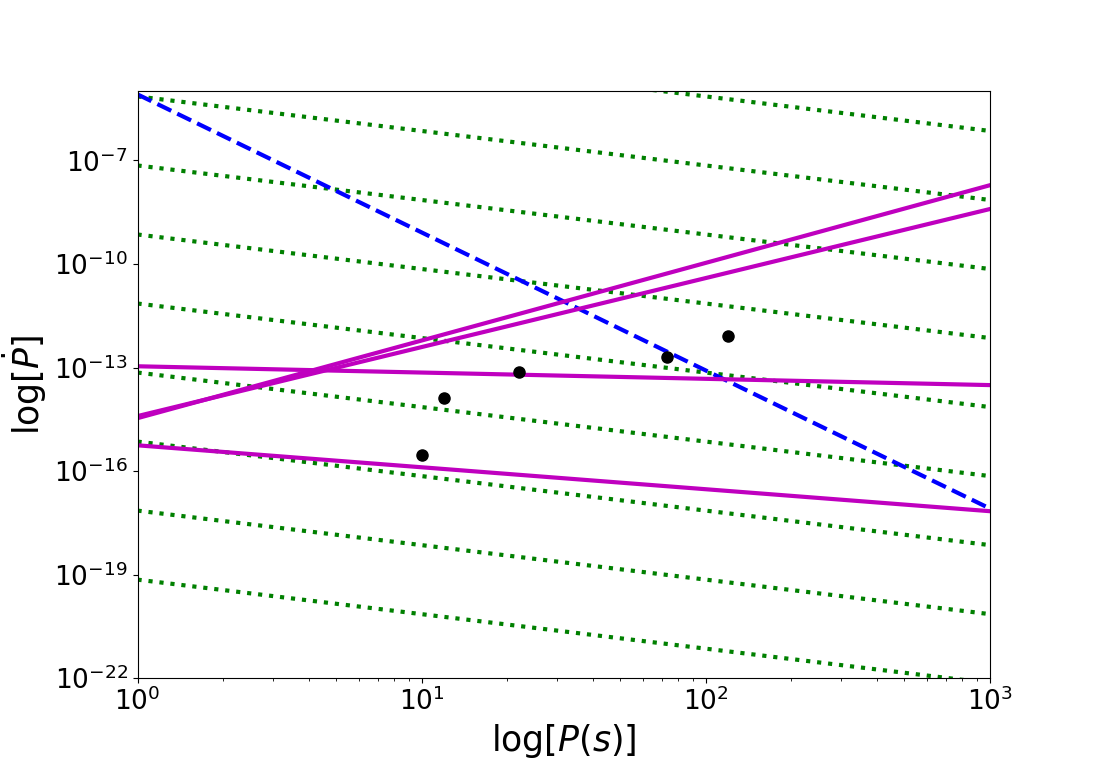}
	\includegraphics[width=7.5cm,page=2]{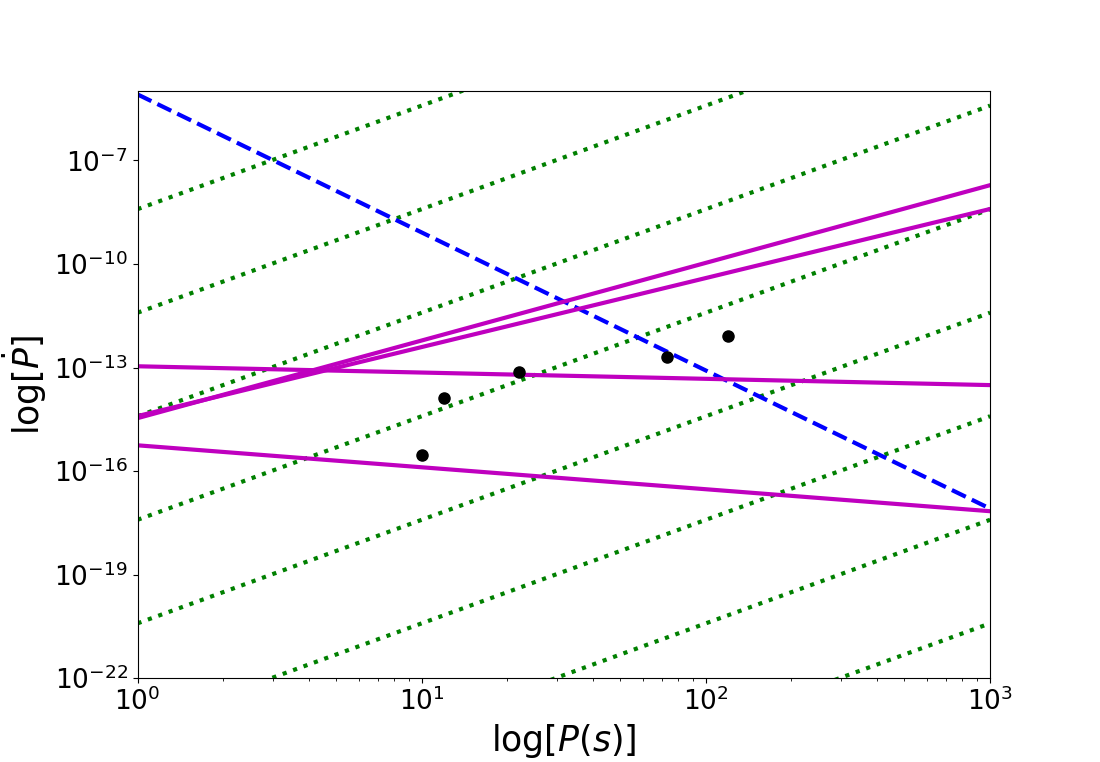}
	\includegraphics[width=7.5cm,page=3]{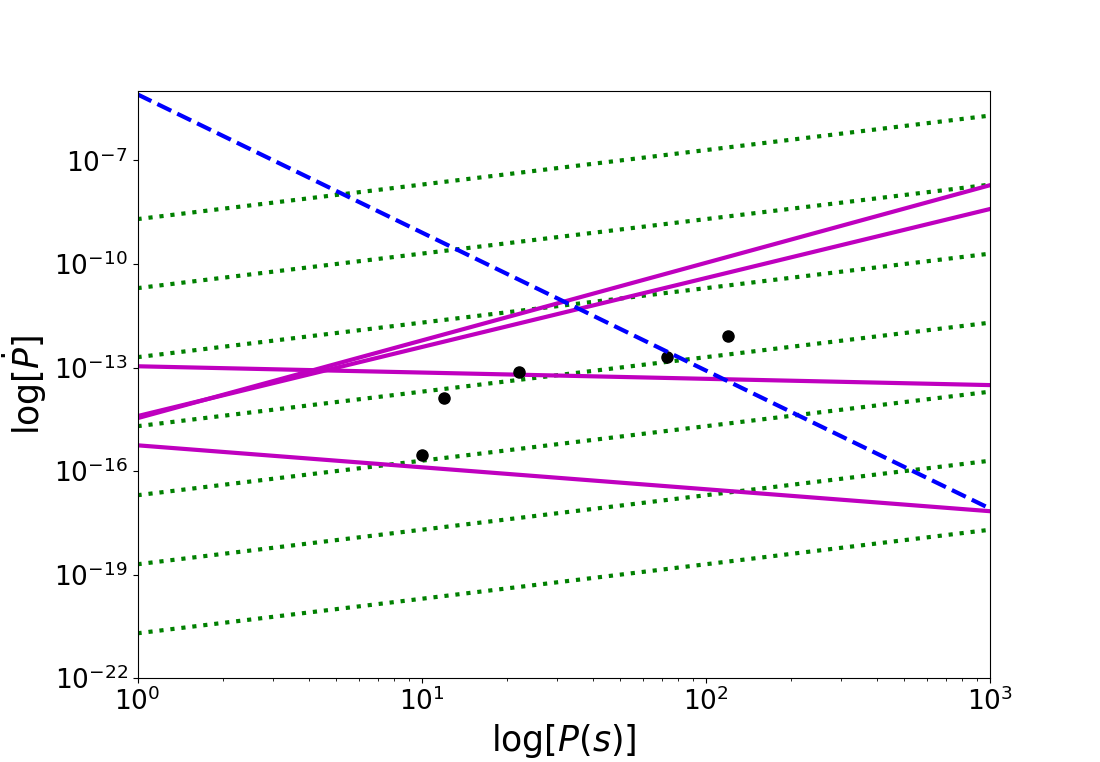}
	\caption{
    Evolution of a white dwarf of mass $1.93M_\odot$, radius $1298.84$ km and
	obliquity angle $\alpha=90^\circ$ with four
	model death lines for constant surface magnetic field (top left panel), constant luminosity (top right panel),
	and constant age (bottom panel). Other parameters are: 
    $n=6$,
	$K=7.6\times 10^{-10}$ CGS, and 
    $K^\prime=2/3c^3$. All lines and points have the same meaning as Fig. \ref{wd90}. The surface magnetic field for PSR~J0901-4046, according to equation (\ref{multipoleB2}), turns out to be $\sim 10^{10}$ G.
    }
    \label{wd9}
\end{figure*}

\begin{figure*}
 	\includegraphics[width=7.5cm,page=1]{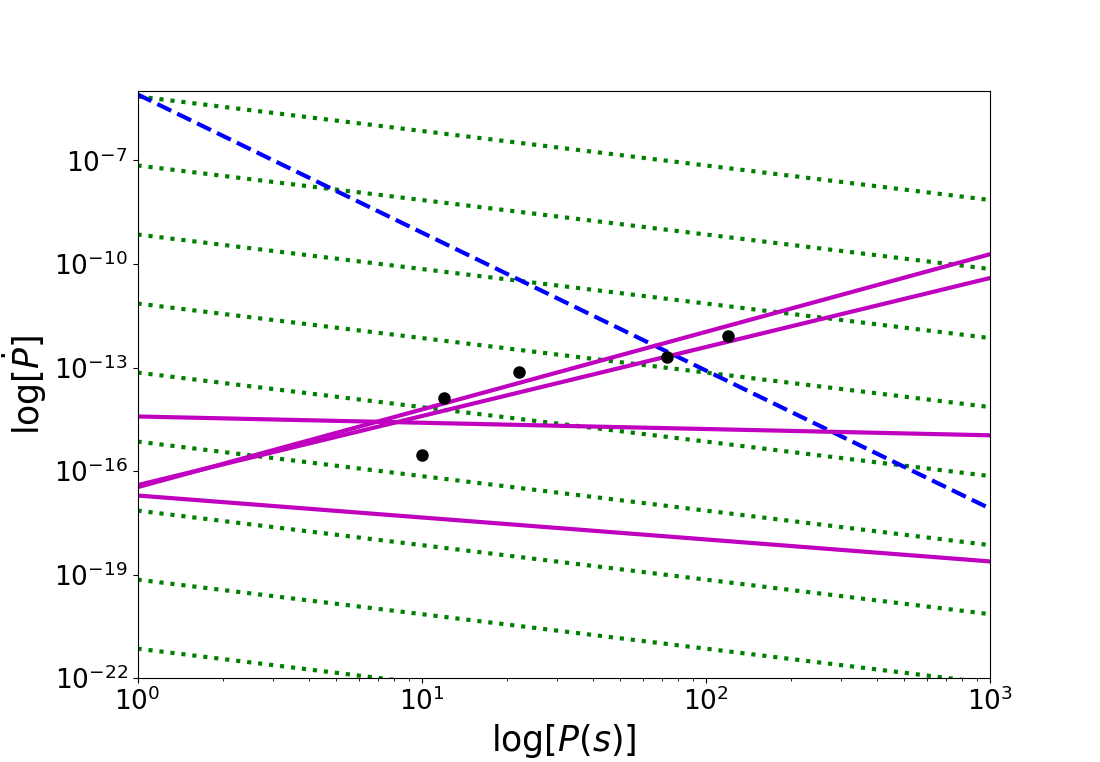}
	\includegraphics[width=7.5cm,page=2]{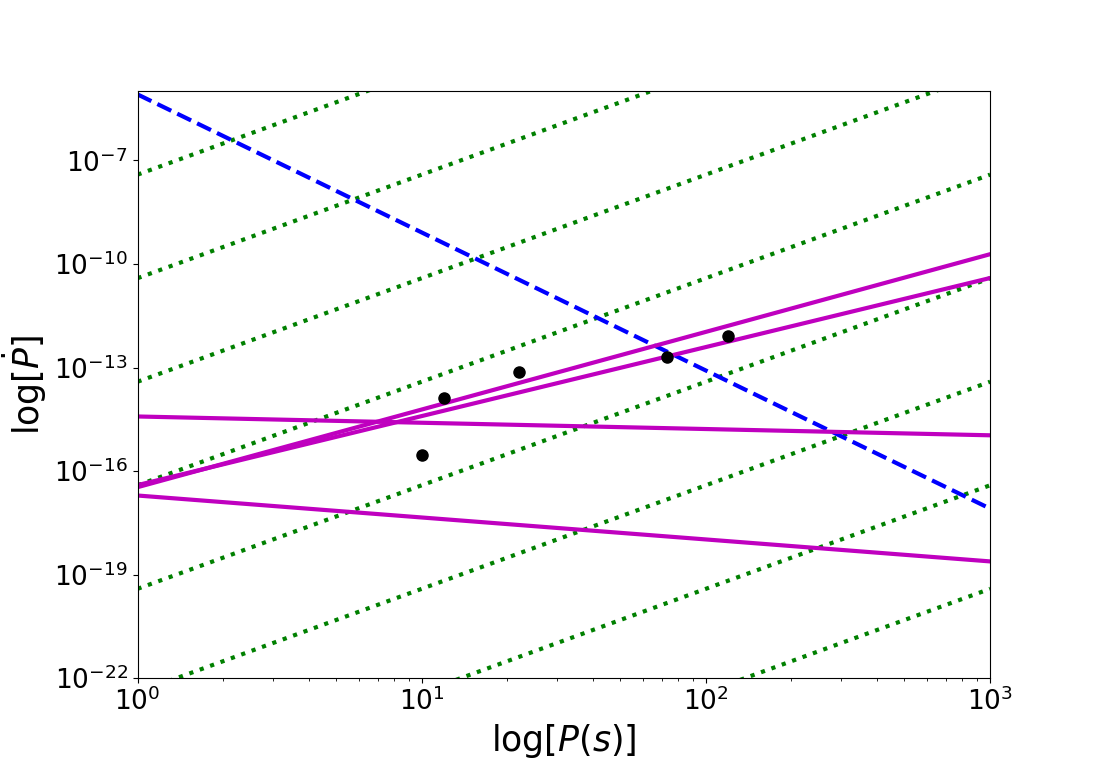}
	\includegraphics[width=7.5cm,page=3]{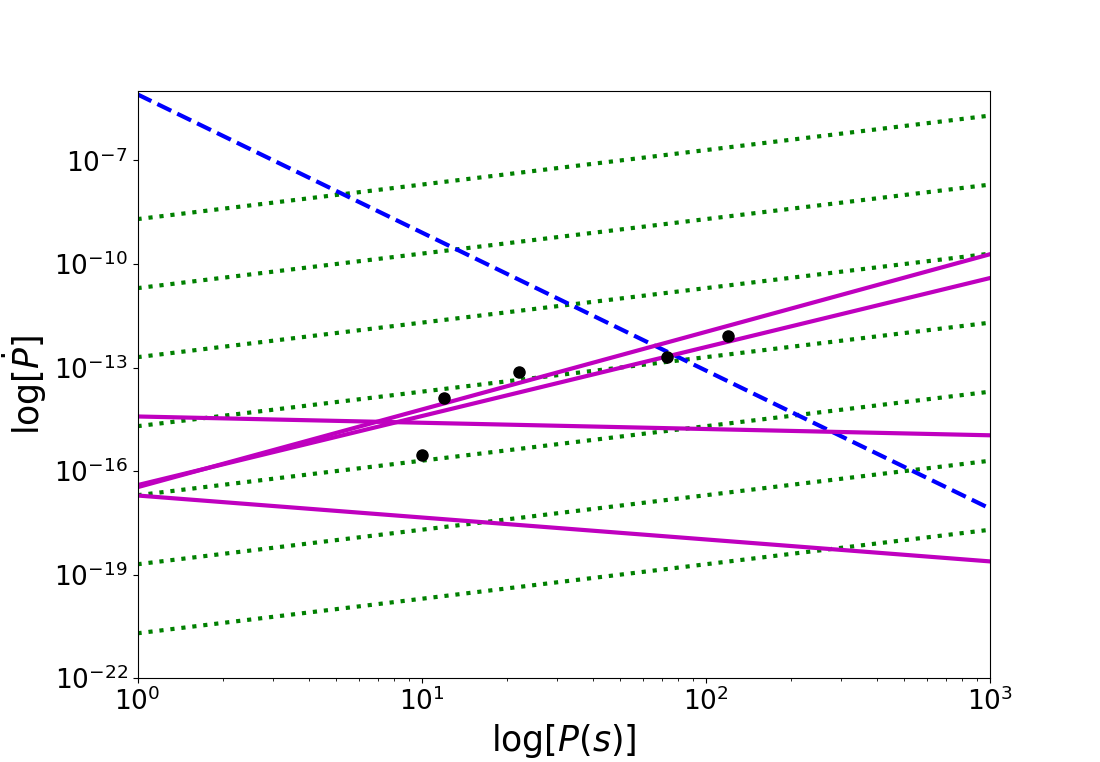}
	\caption{Same as Fig. \ref{wd9}, except for a neutron star of radius $13$ km with the same
	mass and obliquity angle. The various constant surface magnetic field lines (top left panel) from top to bottom correspond to $10^{17}$ to $10^{9}$
G with a common ratio between two successive values 10 and the various constant luminosity lines (top right panel) from top to bottom correspond to $10^{40}$ to $10^{19}$
erg/s with a common ratio between two successive values $10^3$. The constant age lines (bottom panel) are the same as Fig. \ref{wd9}. The surface magnetic field for PSR~J0901-4046, according to equation (\ref{multipoleB2}), turns out to be $\sim 10^{14}$ G. 
}
    \label{ns-p}
\end{figure*}

\begin{figure*}
 	\includegraphics[width=7.5cm,page=1]{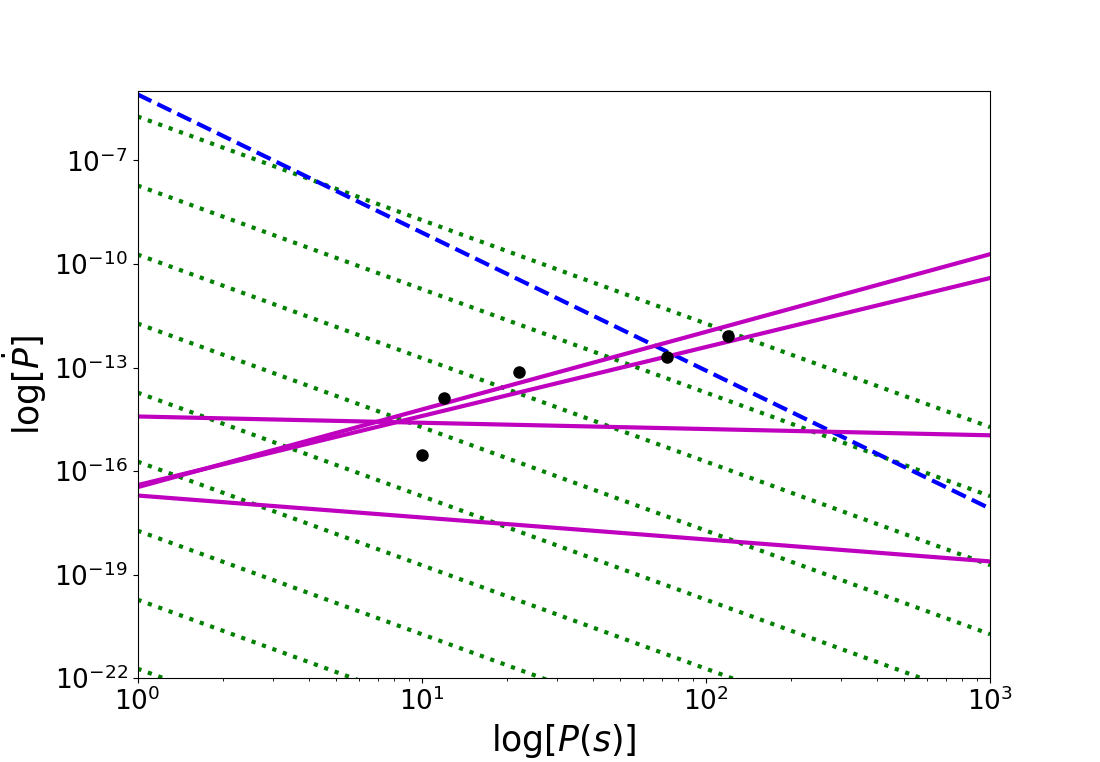}
	\includegraphics[width=7.5cm,page=2]{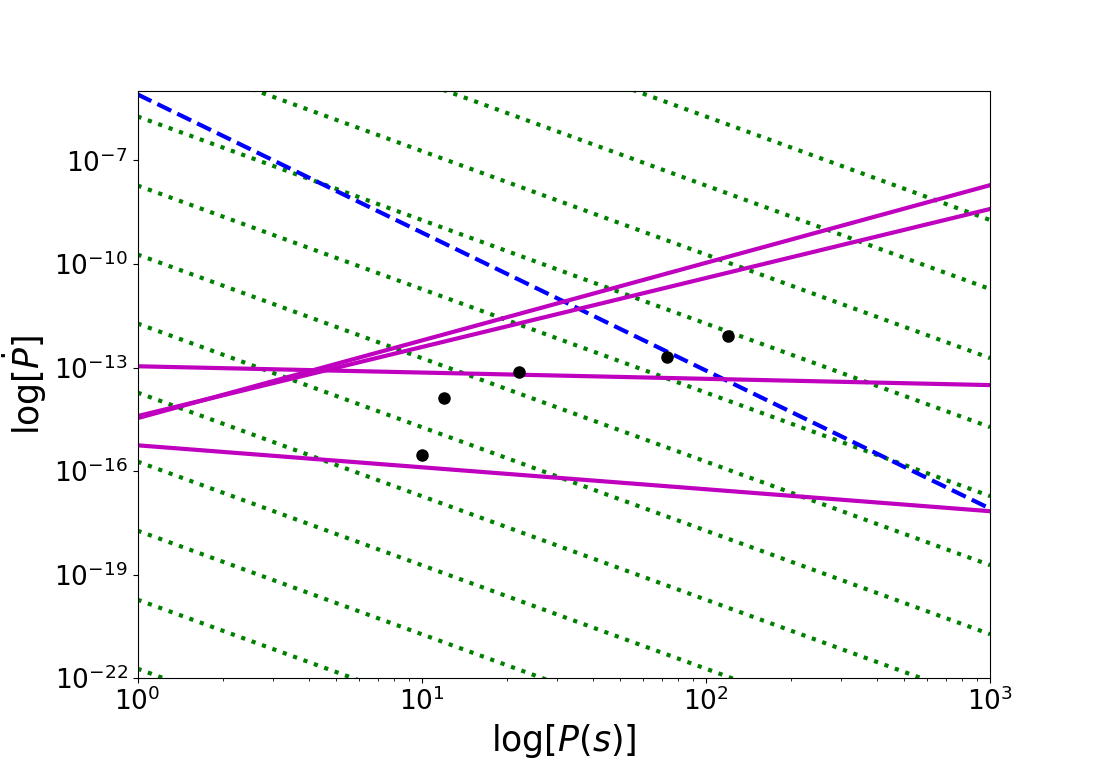}

	\caption{
     Evolution of neutron star (left) and white dwarf (right) of mass $1.93M_\odot$ assuming them leading to quadrupolar emission with their radii of 13 km and 1298.84 km, respectively, and obliquity
	angle $\alpha=90^\circ$ with four
	model death lines for constant surface magnetic field. Other parameters are: $n=6$, $K^\prime=64/25c^5$ \citep{PETRI2015},
	$K=7.6\times 10^{-10}$ CGS. The various constant surface magnetic field lines from
	top to bottom correspond to (left) $10^{20}$ to $10^{13}$ G with a common ratio between two successive values 10 for neutron star, and (right) they are $10^{17}$ to $10^6$ G with a common ratio 10 for white dwarf.  
	The filled circles are the same as in Fig. \ref{wd90}. 
    The dashed line passing 
	through J0901-4046 corresponds to equation (\ref{omn}).
}
    \label{quadNW}
\end{figure*}

Table \ref{table:STARS_WDP} suggests that a white dwarf hypothesis explains PSR~J0901-4046 with much lower magnetic fields, however with lower obliquity angles. We find that the orthogonal rotator ($\alpha = 90\degree$) model does not give an exact match with the values obtained from the stellar evolution simulation, unlike the theoretical exploration above. Further, $B_s$ of the simulation results match with theory only for a bigger, possibly toroidally dominated \citep{SSM2015} B-WD. In fact, simulations favor massive B-WDs, if $B_s$ has to be higher, with lower obliquity angle, corroborating with other studies, e.g. based on duty cycle \citep{J0901_10deg}. By a simple extrapolation, the magnetic fields corresponding to Figs. \ref{wd90}, \ref{wd9} and \ref{ns-p}, and Table \ref{table:field} would scale with $(\sin\alpha)^{-1}$ for various $\alpha$. See the small $\alpha$ theoretical cases reported in Table \ref{table:field}, approximately corroborated with the simulated results reported in Table \ref{table:STARS_WDP}.

\begin{table}[!h]
\centering
\begin{tabular}{lllll}
\hline
        $M/M_\odot$ & $R/\rm{km}$ & $\alpha$    & $B_s$/G            \\ \hline
  0.75         & 12495      & $6\degree$ & $6.41 \times 10^8$ \\
  1.35       & 8295        & $8\degree$ & $1.45 \times 10^9$ \\
  2.04       & 4355        & $8\degree$ & $5.27 \times 10^9$ \\
 2.15       & 4000        & $2\degree$ & $3.13 \times 10^{10}$ \\
\hline
\end{tabular}
\caption{A few of the representative STARS simulated white dwarf parameters where the magnetic field was sufficient to explain the corresponding $P$ and $\dot{P}$ for J0901-4046.}
\label{table:STARS_WDP}
\end{table}

\section{Summary and Conclusions}

We show that PSR J0901-4046 fits well with the machinery of B-WD and, in general, magnetized white dwarf. Independently, whether the underlying white dwarf is sub-Chandrasekhar (but magnetized without affecting, however, Chandrasekhar's mass--radius relation) or super-Chandrasekhar, magnetized white dwarfs can abide with the properties of the source. Further, STARS simulations suggest it to be super-Chandrasekhar, if the surface field is $\sim 10^{10}$ G with orders of magnitude higher interior fields, as theory suggests. 
It is consistent with the observed 
luminosity of the source by its spin-down energy release rate.
The inferred magnetic field is also not unusually high and is close to the magnetic fields observationally inferred for the other white dwarfs. 
This also helps constrain the models of the death line. 

On the contrary, while the unusually slowly rotating neutron star hypothesis can explain some properties of PSR J0901-4046, the underlying magnetic field needs to be unusually high, and could even be an order of magnitude higher than the fields generally assumed to be in magnetars \citep{J0901_10deg}, unless the emission is by ECME which is not free from caveats. Hence, there is a good possibility that PSR~J0901-4046 is a white dwarf or B-WD.
It is now to be checked by other observations, e.g. polarization, of the source to conform with a white dwarf. The bigger size of a white dwarf better fits with the idea of multipolar components including magnetospheric effects along with toroidal field, which needs to be explored in detail. Measurements of linear and circular polarizations and polarization position angle, based on Stokes parameters, reveal several fundamental characteristics, like field strength and geometry, of the radio emitting sources. This helps in verifying them to be B-WDs. 
Future polarization modeling and deeper multi-wavelength observations will be crucial to distinguish definitively between the B-WD and neutron star scenarios.

One of the critical differences between B-WDs and neutron stars as the possible compact object sources for such peculiar, slowly rotating pulsars is their size. Neutron stars have radii $\sim 10$ km, whereas B-WDs can have much larger radii, a few thousand km (see Fig. \ref{evol-acc}). Hence, any size indicator, like the size of the thermal blackbody radiation, should differ substantially in these two classes of compact objects. The sizes inferred during their high-intensity states probably could indicate the size of the local hot spots. AR Sco, known to be a white dwarf, shows thermal components at the characteristic temperatures of 0.1 keV and 1.8 keV which, if interpreted as blackbody radiations, infers an unreasonably small size of 0.1 km \citep{takata} for emitting region. The emissions during the quiescent states, however, could be a more accurate indicator for the sizes of the compact objects. The X-ray emissions from pulsars and magnetars were examined \citep{chu23} in quiescence and it was found that the thermal emission sizes are a few km (the hotspots, however, could be as small as 0.02 km). Attempts to measure the quiescent state fluxes from even the closest isolated radio pulsar, PSR~J2144–3933, could only provide upper limits in soft X-rays \citep{tiengo}. More sensitive measurements in X-rays and UV of such slowly rotating pulsars could perhaps yield some indicators of the sizes of the compact objects in these sources. The larger sizes of B-WDs provide a more extensive reservoir of the rotational energy and, hence, we can also speculate that any local relationship between $\dot{P}$ and luminosity should differ substantially between neutron stars and B-WDs. 

STARS simulations suggest that the temperature of a B-WD or highly magnetized white dwarf, like proposed
PSR~J0901-4046, is $\lesssim 10^4$ K. 
Hence, from Table \ref{table:STARS_WDP}, the UV luminosity turns out to be
$L_{\rm UV}\lesssim 10^{30}$ erg s$^{-1}$ if the entire star emits. However, the emission region may be very small in the stellar surface, as mentioned above. This suggests that $L_{\rm UV}$ could be as small as $10^{20}$ erg s$^{-1}$, which is in accordance with no UV observation yet from PSR~J0901-4046. 
Note that there may be localized regions where the surface temperature is high enough for non-resonant ICS to happen. This is particularly expected in the presence of multipolar field structures leading to the formation of hotspots and subsequent magnetic reconnection \citep{GilZhang05}. This, along with no X-ray observation, is expected to be the possible reason for such a source to be observationally rare. To detect them, one may need to rely upon other features, like, radio emission from a strongly magnetized large star as a B-WD is. As B-WDs seem to be an unusual class of white dwarf, a source like PSR~J0901-4046 is rare.

Hence, it appears that some rotation-powered pulsars exhibiting 
extreme observed properties could actually be white dwarfs rather than neutron stars. It is demonstrated that the B-WD hypothesis alleviates the requirement for an unusually high magnetic field under the neutron star hypothesis. 
In Paper II, we explore other unusual sources under the B-WD framework to see if they fit better. We have already briefly touched upon GLEAM-X J162759.5-523504.3 in the present work, which shows promise to be a B-WD.

\section*{Acknowledgements}
The authors thank Pawan Kumar of UT Austin for discussion. Thanks are also due to the referee for encouraging comments and suggestions to improve the presentation.
BM acknowledges partial support by a project of the SERB, India with research Grant No. CRG/2022/00346. BM further thanks the Institute of Astronomy, University of Cambridge, for hospitality
where part of the work was done. ZZ acknowledges the Prime Minister's Research Fellows (PMRF)
scheme, with Ref. No. TF/PMRF-22-7307, for providing fellowship.

\section*{Data Availability}

The data used in this study are available from the authors upon reasonable request.



\bibliography{draft_m3} 
\bibliographystyle{aasjournalv7}

\label{lastpage}
\end{document}